\documentclass[aps,prl,twocolumn,groupedaddress,showkeys,showpacs,floatfix,superscriptaddress]{revtex4-1}
\usepackage{graphicx}
\usepackage{amsmath,amssymb}
\usepackage{enumerate}
\usepackage{color}
\usepackage{bm}
\usepackage[dvipdfm,%
bookmarks=true,%
bookmarksnumbered=true,%
bookmarkstype=toc,%
colorlinks=true,%
linkcolor=blue,%
citecolor=blue,%
pdfauthor={Naofumi Tsunoda},%
pdftitle={sdpf},%
pdfkeywords={effective interaction,nondegenerate}%
]{hyperref}

\def\~#1{\tilde{#1}}
\def\Nu#1#2#3{{}^{#2}_{#3}\mathrm{#1}}

\newcommand\Heff{H_{\mathrm{eff}}}
\newcommand\HBH{H_{\mathrm{BH}}}

\newcommand\Vlowk{V_{\mathrm{low}k}}

\newcommand\fmi{\mathrm{fm}^{-1}}

\newcommand\diff{\mathrm{d}}

\newcommand\Qbox{$\hat{Q}$-box }

\newcommand\ntlo{\chi\mathrm{N}^3\mathrm{LO} }
\newcommand\hw{\hbar\omega}

\begin{document}

\bibliographystyle{apsrev}

\title{Exotic neutron-rich medium-mass nuclei with realistic nuclear forces}

\author{Naofumi Tsunoda}
\affiliation{Center for Nuclear study, the University of Tokyo, 7-3-1
Hongo, Bunkyo-ku, Tokyo, Japan}
\author{Takaharu Otsuka}
\affiliation{Center for Nuclear study, the University of Tokyo, 7-3-1
Hongo, Bunkyo-ku, Tokyo, Japan}
\affiliation{Department of physics and Center for Nuclear Study,  
the University of Tokyo, 7-3-1 Hongo, Bunkyo-ku, Tokyo, Japan}
\affiliation{National Superconducting Cyclotron Laboratory,
Michigan State University, East Lansing, MI, 48824, USA}
\affiliation{Instituut voor Kern- en Stralingsfysica, Katholieke Universiteit Leuven, B-3001 Leuven, Belgium}
\author{Noritaka Shimizu}
\affiliation{Center for Nuclear study, the University of Tokyo, 7-3-1
Hongo, Bunkyo-ku, Tokyo, Japan}
\author{Morten Hjorth-Jensen}
\affiliation{National Superconducting Cyclotron Laboratory and Department of Physics and Astronomy,
Michigan State University, East Lansing, MI, 48824, USA}
\affiliation{Department of Physics, University of Oslo, N-0316 Oslo, Norway}
\author{Kazuo Takayanagi}
\affiliation{Department of Physics, Sophia University, 7-1 Kioi-cho,
Chiyoda-ku, Tokyo 102, Japan}
\author{Toshio Suzuki}
\affiliation{Department of Physics, College of Humanities and Sciences,
Nihon University, Sakurajosui 3, Setagaya-ku, Tokyo 156-8550, Japan}

\date{\today}

 \begin{abstract}
  We present the first application of the newly developed EKK theory of the 
  effective nucleon-nucleon interaction to shell-model studies of exotic nuclei, including those 
  where conventional approaches with fitted interactions encounter difficulties.
  This EKK theory enables us to derive the interaction suitable for
  several major shells ($sd$+$pf$ in this work). 
  By using such an effective interaction obtained from the Entem-Machleidt QCD-based $\ntlo$ interaction and 
  the Fujita-Miyazawa three-body force, 
  the energies, E2 properties and spectroscopic factors of low-lying states of 
  neutron-rich Ne, Mg and Si isotopes are nicely described, 
  as the first shell-model description
 of the ``island of inversion'' without fit of the interaction.      
  The long-standing question as to how particle-hole excitations occur across the $sd$-$pf$ 
  magic gap is clarified with distinct differences from the conventional approaches.
  The shell evolution is shown to appear similarly to earlier studies.
\end{abstract}

 
\maketitle

{\it Introduction.} -- The nuclear shell model \cite{Mayer:1949,Jensen:1949} provides a
unified and successful description 
of both stable and exotic nuclei, as a many-body framework which can be related 
directly to nuclear forces.  Exotic nuclei are located far from 
the $\beta$-stability line on the Segr\`e chart, exhibiting very short life times, mainly due to an
unbalanced ratio of proton ($Z$) and neutron ($N$) numbers.  Exotic
nuclei differ remarkably in some other aspects from their stable counterparts, 
providing us with new insights in understanding atomic nuclei and nuclear forces 
~\cite{Sorlin:2008er,Gade:2008we,nobel_otsuka}.  As experimental data
on exotic nuclei are, in general, less abundant compared to stable
nuclei, theoretical calculations, interpretations and predictions play
an ever increasing role.

Shell-model (SM) calculations handle the
nuclear forces in terms of two-body matrix elements (TBMEs).  
In the early days, TBMEs were empirically
determined in order to reproduce certain observables.  
A well-known example is the effective interaction for
$p$-shell nuclei by Cohen and Kurath~\cite{Cohen:1965kz}.  A
breakthrough towards more microscopically-derived TBMEs was
achieved by Kuo and Brown for $sd$-shell nuclei~\cite{KuoBrown}.
Although basic features of the nucleon-nucleon ($NN$) force for the SM calculation 
are included in these effective interactions, empirical adjustments of TBMEs
were needed in order to reproduce various observables 
\cite{Brown1988191,RevModPhys.77.427,PhysRevC.65.061301}.

These effective interactions were all derived for a Hilbert
space represented by the degrees of freedom of one major (oscillator) 
shell. As we move towards exotic nuclei,
some new features and phenomena arise.
A notable example can be the shell evolution due to
nuclear forces between a proton and a neutron in different shells 
\cite{PhysRevLett.95.232502,Otsuka:2009qs,nobel_otsuka}.
This leads to significant particle-hole excitations between two shells, for example,  
in $Z$=8-14 neutron-rich exotic nuclei  
\cite{PhysRevLett.95.232502,Otsuka:2009qs,nobel_otsuka,Sorlin:2008er,Gade:2008we,Campi:1975un,Warburton:1990vw,Fukunishi:1992ic,PhysRevC.60.054315,Caurier:2013aoa,RevModPhys.77.427}. 
A microscopic understanding of many of these phenomena requires the degrees of
freedom of at least two major shells. 

To derive SM effective Hamiltonians 
is a challenge to nuclear theory. Several attempts have
been made recently in this direction
\cite{LisetskiyPhysRevC78,Bogner.PhysRevLett113,PhysRevLett.113.142502,Coraggio:2013dm,Coraggio:2014cf,Simonis2015,Jansen:2016ik},
while the issue of two major shells is still unsettled. 

The aim of this work is first to derive SM interaction for the model
space consisting of the $sd$ and $pf$ shells 
based on the so-called Extended Kuo-Krenciglowa
(EKK) method \cite{Takayanagi201161,Takayanagi:2011dv,Tsunoda:2014hj}.
Secondly, we apply this interaction to our studies of exotic neutron-rich Ne, Mg and Si isotopes. These are nuclei in and around
the so-called ``island of inversion'' \cite{Warburton:1990vw}, 
where the degrees of freedom of $sd$ and $pf$ shells are essential.
We thus present, in this Letter, the first application of the EKK method to actual cases.
Three-nucleon forces (3NFs) are also included 
since they play an important role in
reproducing basic nuclear properties  
~\cite{PhysRevLett.105.032501,Holt:2012gc,hagen2012a,hagen2012b,Ekstrom:2015gw}.

{\it Hamiltonian and model spaces.} --
Many-body perturbation theory (MBPT) has been the
method of choice for deriving effective interactions for the nuclear shell model, see for example
Refs.~\cite{Kuo:1971em,Krenciglowa1974171, Krenciglowa1977381,Kuo_springer,HjorthJensen1995125}.  
The conventional MBPTs, for instance, Kuo-Krenciglowa (KK) method 
~\cite{Kuo:1971em,Krenciglowa1974171, Krenciglowa1977381}, are constructed 
for degenerate single-particle states in the model space, which usually refers to one major shell
~\cite{Kuo:1971em,Krenciglowa1974171, Krenciglowa1977381,Kuo_springer,HjorthJensen1995125}.   
The present model space, however, includes all single-particle states of the
$sd$ and the $pf$ shells, labeled $sdpf$ hereafter.  
This leads to possible divergences when
constructing an effective interaction with conventional MBPTs due to the
non-degeneracies of the 
single-particle states.
A practical, but not satisfactory way to
circumvent such divergences has been to enforce in an {\em ad hoc} way degenerate
single-particle energies for all involved single-particle
states spanning a model space, see for example
Refs.~\cite{Holt:2012gc,Holt:2013en,Holt:2014uc}.  
Employing the EKK method as done in this work, removes the
above-mentioned divergence problems and allows for a correct treatment of
the different single-particle energies \cite{Tsunoda:2014hj}.
Starting from a Hamiltonian which contains kinetic energy and a two-body $NN$ interaction and
following the formalism described in Ref.~\cite{Tsunoda:2014hj}, we can define  an effective Hamiltonian through
\begin{equation}\label{eq:EKK}
\Heff=\HBH(\xi)+
 \sum_{k=1}^{\infty}\frac{1}{k!}\frac{\diff^k \hat{Q}(\xi)}{\diff \xi^k}\{\Heff -\xi\}^k, 
\end{equation}
where $\HBH$ and $\hat{Q}$ are the Bloch-Horowitz Hamiltonian and the so-called \Qbox \cite{Kuo_springer,HjorthJensen1995125}, respectively. 
The latter contains only linked and unfolded  
Feynman-Goldstone diagrams in MBPT.
The quantity $\xi$ is a parameter (the origin of the \Qbox expansion), and the poles of $\hat{Q}(\xi)$ can be
avoided with appropriate value. 

We renormalize the $NN$ interaction using the so-called $\Vlowk$ approach with a cutoff
$2.0~\fmi$~\cite{Entem:2003hx,PhysRevC.65.051301,PhysRevC.70.061002}.
We solve Eq.~\eqref{eq:EKK} with a  harmonic oscillator basis with the oscillator parameter 
$\hw=45 A^{-1/3} - 25 A^{-2/3}$MeV=12.10 MeV where $A=28$ ($A=Z+N$).  
The \Qbox is calculated up to the third order in Eq. (\ref{eq:EKK}) 
and intermediate excitations up to $17~\hw$
are included to guarantee convergence in the sum over intermediate
states.  Three-nucleon forces are included through the Fujita-Miyazawa
term with its strength given by a standard $\pi$-$N$-$\Delta$ coupling  
\cite{greenRepProgPhys}. 
This term is in turn  transformed into a medium-dependent two-body
interaction, see  
Refs.~\cite{PhysRevLett.105.032501,Holt:2013jk,Holt:2014uc} for details. 
 
The single-particle energies (SPEs) in the $sdpf$ space are input too.  
Because of on-going efforts towards a better description of density
profiles~\cite{Ekstrom:2015gw},
which is directly related to the size of shell gap,
we employ here $^{16}$O as the closed-shell core and fit the values of SPEs to selected 
observables such as ground-state energies of $N<20$, $^{34}$Si 2$_1^+$ level, {\it etc}.  
The energies of the 2p$_{1/2}$ and the 1f$_{5/2}$ orbits are constrained
by the GXPF1 Hamiltonian of Ref.~\cite{PhysRevC.65.061301}, due to lack of 
appropriate data.
These SPEs, {\it i.e.}, one-body part of the Hamiltonian, will be confirmed to be reasonable, 
because effective SPEs to be calculated for $Z$=$N$=20 will turn out close to SPEs known for $^{40}$Ca.  
The SPEs are isospin invariant, because of the $^{16}$O core.

In this work we study Ne ($Z$=10), Mg ($Z$=12) and Si ($Z$=14) isotopes
with a focus on $N\sim20$ nuclei.
Eigenvalues and eigenstates are obtained via standard Lanczos diagonalization method, 
including up to 8-particle-8-hole (8p8h) 
excitations from the $sd$-shell to the $pf$-shell. 
For the cases where the Lanczos diagonalization is infeasible ({\it e.g.}, $N>22$) because of 
too large dimension, we switch to a Monte Carlo Shell Model calculation \cite{Otsuka:2001gx,Shimizu:2012dh}. 

The TBMEs vary gradually as a function of $A$ because $\hw$ changes.  
Although TBMEs can be calculated for each nucleus explicitly,  
we introduce an overall scaling factor $(A/A_0)^{-0.2}$ with  
$A_0=28$.  The value of the power is determined so that TBMEs calculated
explicitly for several nuclei can be best reproduced by this simple parametrization. 

\begin{figure}[tbh]
 \includegraphics[width=8cm,angle=0,clip]{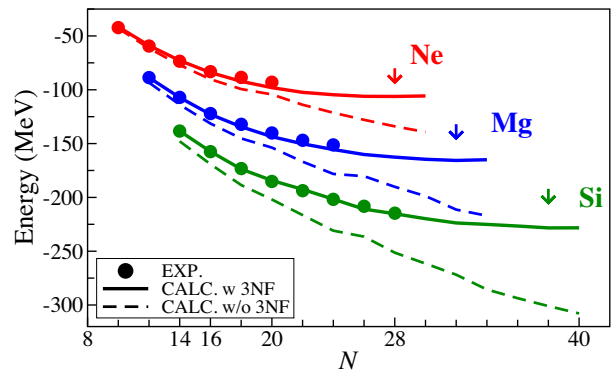}
 \caption{
 Ground-state energies of Ne, Mg and Si isotopes. Circles represent the experimental
 values.   Shell-model results with (without) 3NF are shown by 
 solid (dashed) lines.
 Arrows indicate the predicted drip lines.
 Experimental data are taken from
 Ref.~\cite{NationalNuclearDataCenter:2008tv}}
 \label{fig:GE}
\end{figure}

{\it Results and discussions.} --
Figure~\ref{fig:GE} shows the experimental ground-state energies
(Coulomb-force effects subtracted) of Ne, Mg and Si isotopes compared
with our theoretical calculations with and without three-body forces.

Figure~\ref{fig:GE} indicates that the ground-state energies are well
reproduced  when 3NF effects are included.
It is also worth noticing that the repulsion due to the 3NF
grows as $N$ increases in all the isotopic chains. This is 
consistent with earlier studies~\cite{PhysRevLett.105.032501,Holt:2014uc}.
We point out that this repulsion becomes stronger also as $Z$ increases,
suggesting that the present 3NF contribution is repulsive also in the proton-neutron 
channel. 
Figure~\ref{fig:GE} shows also present predictions for the drip lines~\cite{koura2005,Erler:2012cc}.

\begin{figure*}[tbp]
 \includegraphics[width=15cm,angle=0,clip]{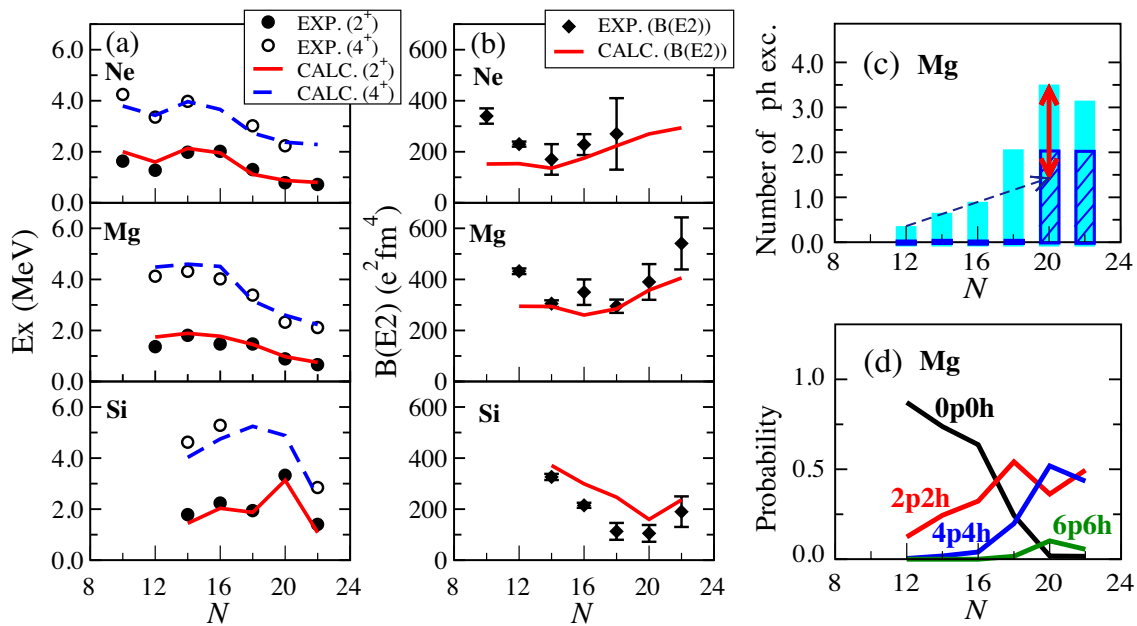}
 \caption{
 (a) Excitation energies of first $2^{+}$
  and $4^{+}$ states and (b) $\mathrm{B(E2}: 0_1^{+}\rightarrow 2_1^{+}$) values 
  of Ne, Mg and Si isotopes.
 Experimental data (symbols) ~\cite{NationalNuclearDataCenter:2008tv,Takeuchi:2012ch} 
 and present calculations (lines) are compared.
 (c) Expectation values of the number of the particle-hole excitations in the ground state of Mg isotopes.
   The plain histograms are present results, while the hatched ones imply the value 
   by \cite{Warburton:1990vw}.  The dashed line shows the trend.
   The two-way arrow indicates the additional 2p2h excitation (see the text).  
   (d) Decomposition of ph-excitation probabilities for the ground state of Mg isotopes. }
  \label{fig:E2_BE2}
\end{figure*}

Figure~\ref{fig:E2_BE2} (a) shows the $2^{+}$ and $4^{+}$ energy levels.
Experimental levels are well reproduced by the present calculation, including 
both gradual and steep changes as a function of $N$.
 For instance, the 2$^+_1$ level drops down steeply
by $\sim$1 MeV from $N=16$ to 18 in Ne chain, while a similar 
change occurs from $N=18$ to 20 in Mg chain.  
The $4^{+}_1$ levels show a different behavior as functions of $N$.
For instance, in Mg isotopes, from $N=16$ to 18, it starts to come down whereas
the $2^{+}_1$ level stays constant, implying that 
not only the degree of deformation but also the shape is changing.
On the other hand, in Si isotopes, the 2$^+_1$ level jumps up at $N=20$.  
These behaviors of excited levels are reproduced well 
by the present calculation.  
The low excitation energy of the $2^{+}_1$ levels in $\Nu{Ne}{30}{}$ and 
$\Nu{Mg}{32}{}$ indicates that these nuclei are strongly deformed and that 
$sd$-to-$pf$ particle-hole (ph) excitations over the $N=20$ gap occur significantly.  
In contrast, the higher-lying $2^{+}_1$ level of $\Nu{Si}{34}{}$
suggests weaker degrees of ph excitations.

Figure~\ref{fig:E2_BE2} (b) shows 
$\mathrm{B(E2}: 0_1^{+}\rightarrow 2_1^{+}$) values. 
With effective charges, $(e_p,e_n)=(1.25, 0.25)e$,  
calculated B(E2) values exhibit systematic behaviors in agreement with experiment. 
This agreement is, however, not as good as that obtained for the energy levels.  
This probably indicates the need to derive E2 operator using the same microscopic theory.
While the $N=20$ shell closure in Si isotopes is evident also in B(E2) systematics, 
the B(E2) values of Ne and Mg isotopes are larger at $N=20$ 
than at $N=18$, a feature which is consistent with growing deformation 
seen in the $2^{+}_1$ level.

\begin{figure}[tbp]
  \includegraphics[width=6.5cm,angle=0,clip]{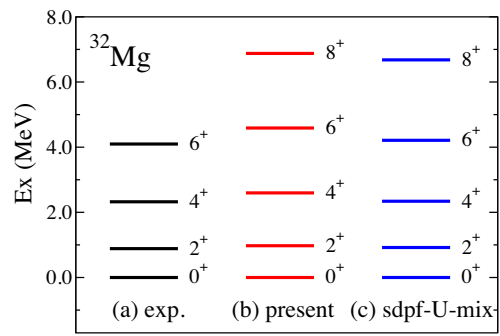}
 \caption{Yrast levels of $\Nu{Mg}{32}{}$ obtained by (a) experiment \cite{Crawford:2016cd},
 (b) present work, and (c) calculation with the sdpf-U-mix interaction \cite{Caurier:2013aoa,Crawford:2016cd}.}
 \label{fig:32Mg_high}
 \end{figure}

Regarding higher spin states, Fig.~\ref{fig:32Mg_high} shows, as an example, 
yrast levels of $\Nu{Mg}{32}{}$, of which the 
6$^{+}$ level was measured recently by Crawford {\it et al.}  \cite{Crawford:2016cd}.  
The present result is in agreement with observed levels as well as those calculated 
by the sdpf-U-mix interaction  \cite{Caurier:2013aoa}.
The calculated B(E2;$J^+ \rightarrow (J-2)^+$) values are 70, 96, 36, and 58, (e$^2$fm$^4$) 
for $J$=2, 4, 6, and 8, respectively.  We note that the 6$^{+}$ and 8$^{+}$ members of the yrast 
band are mixed with less deformed states in the present calculation, 
reducing their energies and B(E2) values from the rotor systematics.
 
We now turn to properties of the wave functions of Mg isotopes.  Figure~\ref{fig:E2_BE2} (c) 
depicts the expectation value of the number of ph excitations over the $Z$=$N$=20 
shell gap for the ground state.  One notices the abrupt increase 
at $N$=18 and furthermore at $N$=20.  These increases are associated with the onset of large deformation.
More particles (predominantly neutrons) in the $pf$-shell and many holes in the $sd$-shell enhance   
collective motion towards more deformed nuclear shapes.
The average value of the ph excitations is about 3.5 for $\Nu{Mg}{32}{}$.  This is a large number, 
compared to the conventional picture for $\Nu{Mg}{32}{}$ being basically a 2p2h state 
\cite{Warburton:1990vw,Fukunishi:1992ic,PhysRevC.60.054315,Caurier:2013aoa,RevModPhys.77.427}.  
The value assumed in Warburton-Brown-Becker's island-of-inversion model \cite{Warburton:1990vw} 
is shown in Fig.~\ref{fig:E2_BE2} (c) as a reference.  The basic trend remains unchanged in other shell-model 
calculations, although the actual values can be somewhat larger \cite{Fukunishi:1992ic,PhysRevC.60.054315,Caurier:2013aoa,RevModPhys.77.427}.  

In the present calculation, this ph value starts low ($<$0.5) at $N$=12, and increases almost 
monotonically to $N$=16.  
This increase is a ``modest'' effect of the effective interaction shifting nucleons between the two shells, 
{\it e.g.} pairing effects. 
Considering that the 2$^+_1$ level remains high ($\sim$ 2 MeV) up to $N$=18, 
this ph excitation mode does not produce a strong deformation.
We extrapolate it linearly up to $N$=20 in Fig.~\ref{fig:E2_BE2} (c).  
The difference between the extrapolated and actual values is about 2, which can be
interpreted as an additional 2p2h excitation essential for promoting the strong deformation.
Although this interpretation is intuitive, the 2p2h excitation on top of the ``modest'' correlation 
appears to be analogous to the 2p2h picture seen in the conventional approaches.

Figure~\ref{fig:E2_BE2} (d) shows more details of the ph excitations involved in the Mg 
ground states.  The probability of the 0p0h configuration comes down slowly for $N\le 16$, and drops down
sharply after that.  The 2p2h probability increases gradually until $N$=16.  The 4p4h is negligible 
up to $N$=16, but increases abruptly for $N \ge 18$, especially for $N$=20.  
Note that the 2p2h probability even decreases at $N$=20.  
Such changes drive the nucleus towards stronger deformation  
by including higher ph configurations.   
The present work thus resolves, for the first time, the long-standing question as to how the 
$sd$-to-$pf$ excitation grows from stable to exotic Mg isotopes,
showing notable differences from conventional approaches.


In constructing effective $NN$ interactions for the $sd$-shell, {\it e.g.} USD \cite{Brown1988191}, 
certain effects of $sd$-$pf$ ph excitations are renormalized  
into TBMEs within the $sd$ shell.  
If an $sd$-$pf$ interaction is constructed on top of this $sd$-shell
interaction (with minor changes),
a renormalized fraction of those excitations should not appear explicitly.    
In fact, this is the case with the $sdpf$-$m$ interaction (probably as well as others), 
with the above-mentioned expectation value 
being as small as 0.05, 0.08, 0.20, 0.85, 2.25 and 1.96 for $^{24-34}$Mg, respectively.
In some cases, however, the explicit treatment of those excitations becomes crucial.

\begin{figure*}[tbp]
 \includegraphics[width=14cm,angle=0,clip]{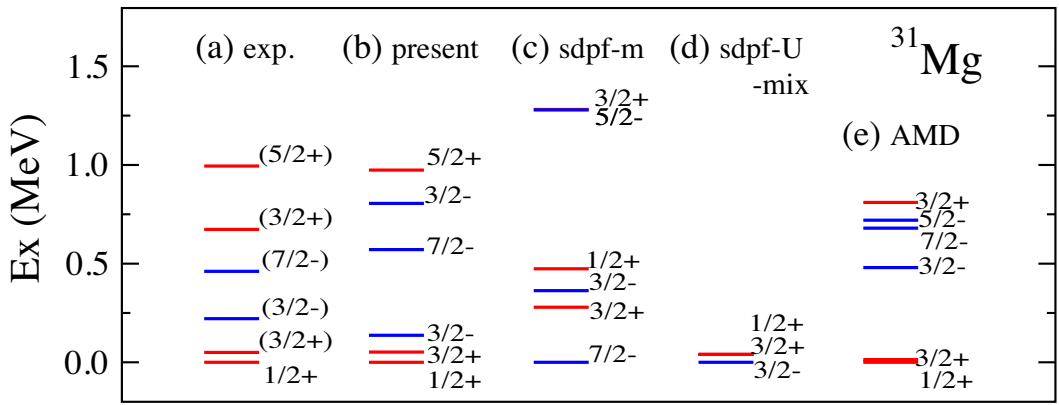}
 \caption{Energy levels of $\Nu{Mg}{31}{}$. (a) experimental values, (b)
 present work, (c) sdpf-m~\cite{PhysRevC.60.054315},
 (d) sdpf-U-mix~\cite{Caurier:2013aoa} and (e)
 AMD+GCM calculation~\cite{Kimura:2007id}, respectively.}
 \label{fig:ant}
\end{figure*}

For transitional nuclei around $\Nu{Mg}{32}{}$, the mixing of different ph configurations is indeed crucial
since the deformations contribute less to its binding energy.  
We show low-lying positive- and negative-parity levels of $\Nu{Mg}{31}{}$ in Fig.~\ref{fig:ant}.
The present calculation reproduces the experimental levels rather well,   
for example, the ordering of lowest four levels.
The configurations of the positive-party states are of 0p0h (normal) by 1-3$\%$, 
2p2h by 66-68$\%$, and 4p4h by 29-30$\%$.  
In contrast, the lowest $3/2^{-}$ and $7/2^{-}$ states consist of 
33-35$\%$ 1p1h (normal), 55-57$\%$ 3p3h, and 11$\%$ 5p5h configurations.
Thus, the ph properties differ significantly between positive- and negative-parity states, 
whereas similar ph excitations can be found within the same parity.  
It is clear that the coupling between different ph configurations has to be evaluated precisely.  
Figure~\ref{fig:ant} includes the results of other calculations.  Among them, only 
the shell-model calculation of \cite{Caurier:2013aoa} reproduces the
near-degeneracy of
the three lowest levels. Further fits of TBMEs were not attempted in Ref.~\cite{Caurier:2013aoa}.     
This is natural because of too many parameters with two major shells. The present work however,   
generates all TBMEs from first principles thanks to the EKK method and
no further fine-tuning of the effective interaction is done.

We briefly discuss a test of the present wave functions by transfer reactions.
The spectroscopic factors are calculated as 0.77 and 1.41 for one-neutron
removal from $\Nu{Mg}{32}{}$ ground state to $3/2^{-}_1$ and $7/2^{-}_1$ states, respectively.
The experimental values are 0.59(11) and 1.19(36) \cite{terry}, respectively, which are in good 
agreement after scaling the theoretical values by the usual quenching factor $\sim 0.7$ \cite{gadeSF,gadeSF2}.
Thus, the present wave functions, which look different from those of conventional calculations, 
turn out to be quite consistent with these experimental values.

\begin{figure}[tbp]
  \includegraphics[width=8.5725cm,angle=0,clip]{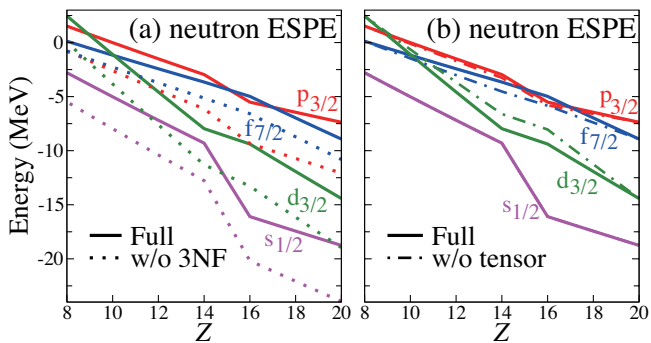}
 \caption{ESPEs of N=20 isotones for neutrons
  obtained in the normal filling scheme.
 Solid (dotted) lines in (a) show the case with (without) three-nucleon forces, while  the  
 solid (dot-dashed) lines in (b) represent the case with (without) the tensor force.}
 \label{fig:ESPE}
 \end{figure}

The last primary subject is the effective single-particle energy (ESPE). The ESPE represents
the combined effect on the single-particle energies from the inert core and the monopole effects 
from other valence nucleons, 
producing important effects on the shell structure of exotic nuclei 
\cite{PhysRevLett.95.232502,Otsuka:2009qs,PhysRevLett.105.032501,nobel_otsuka}.   
Figure~\ref{fig:ESPE} shows the neutron ESPEs of the
$N=20$ isotones, calculated in the normal filling scheme, with and
without  3NF and tensor force component.
Here the tensor force component is extracted from a spin-tensor decomposition
of the present $NN$ effective interaction generated by the EKK method \cite{Tsunoda:2014hj}.
Figure~\ref{fig:ESPE} (a) indicates that the primary effect of 3NF is to raise all orbits by 
similar amounts \cite{PhysRevLett.105.032501}.  
Without 3NF effect, the neutron orbits are bound too deeply, yielding excessive binding energies,  
as seen in Fig.~\ref{fig:GE}.
On the other hand, 
Fig.~\ref{fig:ESPE} (b) demonstrates that the tensor force lowers or raises the ESPEs depending 
on the cases, and thereby 
change rather rapidly the spacing between certain orbits \cite{PhysRevLett.95.232502}.   
The ESPEs obtained at $N=20$, including other $sd-pf$ orbits, appear to be reasonable for $^{40}$Ca, 
suggesting the validity of the input SPEs being used. 
 
Figure~\ref{fig:ESPE} (a,b) indicates that the neutron $sd$-$pf$ gap grows 
rapidly as $Z$ increases
from $Z=8$ to $Z=14$, and stays relatively constant after $Z=14$ up to $Z=20$ 
as being the $N=20$ magic gap.
This gap, however, disappears around $Z=8$, and the $N=16$ magic gap arises instead
~\cite{brownN16,ozawaPRL84,magic}.  
This is consistent with the trend shown by the simplified interaction, called VMU \cite{Otsuka:2009qs}, 
where the central force gives a steady enlargement of the gap from $Z=8$ to $Z=20$, 
while the tensor force makes the change more rapid for $Z < 14$ but cancels the change for  
$Z >14$.
The neutron ESPEs of $f_{7/2}$ and $p_{3/2}$ at $Z=20$ are also consistent
with those from empirically determined interactions~\cite{PhysRevC.60.054315}.
Two other upper orbitals, $p_{1/2}$ and $f_{5/2}$
appear at the energies similar to those of another empirical
interaction~\cite{PhysRevC.65.061301},
although not shown in Fig.~\ref{fig:ESPE}.
We thus demonstrate, for the first time, what type of shell evolution a chiral $NN$ interaction 
such as the $\ntlo$ interaction produces in medium-mass exotic nuclei.   
It is seen also that the lowering as a function of $Z$ is dampened  by three-nucleon forces.
The resulting trend is similar to that obtained  with the VMU
interaction of Ref.~\cite{Otsuka:2009qs} as well as other interactions
fitted to
experiment~\cite{PhysRevC.60.054315,Caurier:2013aoa,RevModPhys.77.427}.


{\it Summary.} --We have presented a consistent microscopic description
of neutron-rich nuclei
$Z$=10-14 and $N\sim$20 employing  the EKK method to derive appropriate multi-shell effective interactions.
The calculated energies, B(E2) values and spectroscopic factors are 
in good agreement with experiment.  Our TBMEs are not fitted and
the quality of the agreement is similar to or better than SM approaches with fitted TBMEs.
Definite differences from conventional approaches are seen, particularly in the 
pattern of particle-hole excitations between the $sd$ and $pf$ shells.  
In this context, the conventional 2p2h picture of the island of inversion 
serves as an intuitive interpretation.
We point out that the above-mentioned differences can be smaller,  
if ph excitations over the relevant magic gap are included explicitly in the fit of TBMEs 
like the case for $pf$+1g$_{9/2}$ orbit, {\it e.g.}, \cite{jun45,68ni}. 
The shell evolution is derived from the chiral EFT force and 
Fujita-Miyazawa 3-body force for the first time.
All these features contribute to further studies of exotic nuclei where microscopic theories 
play crucial roles.  
Further progress in nuclear forces and many-body treatments will improve
the agreement to experiment and provide us with more predictive power.

After the submission of this paper, Macchiavelli {\it et al.} reported~\cite{Macchiavelli:2016cc}
 that the ground state of $\Nu{Mg}{32}{}$ is dominated by 2p2h and 4p4h
 configurations with nearly equal probabilities.   Although this
 result was obtained in an empirical analysis by a one-parameter mixing of 
 three 0$^+$ states and the structure of $\Nu{Mg}{30}{}$ taken in
 \cite{Macchiavelli:2016cc} differs from the present one, 
 it is of interest
 that such large probabilities were obtained in both works independently.
 
\begin{acknowledgments}
We thank Dr. Y.~Utsuno for useful discussions.
The Lanczos shell-model calculation is performed with the code
``KSHELL''~\cite{Shimizu:1613378}.
This work was supported in part by Grants-in-Aid for Scientific Research (23244049,15K05090). 
It was supported in part by HPCI Strategic Program (hp140210,hp150224,hp160221), in part by MEXT and JICFuS 
as a priority issue (Elucidation of the fundamental laws and evolution of the universe) 
to be tackled by using Post ``K'' Computer, 
and also by CNS-RIKEN joint project for large-scale nuclear structure
 calculations.
MHJ acknowledges U.S. NSF Grant No. PHY-1404159 (Michigan State University) and the Research Council of Norway
 under contract ISP-Fysikk/216699.

\end{acknowledgments}

\bibliography{reference}

\end{document}